\documentclass[fleqn,usenatbib]{mnras}
\usepackage{newtxtext,newtxmath}
\usepackage[T1]{fontenc}

\DeclareRobustCommand{\VAN}[3]{#2}
\let\VANthebibliography\thebibliography
\def\thebibliography{\DeclareRobustCommand{\VAN}[3]{##3}\VANthebibliography}

\usepackage{amsmath}
\usepackage{graphicx}
\usepackage{enumerate}   
\usepackage{color,xcolor}
\hypersetup{colorlinks=true,citecolor=blue,urlcolor=blue,linkcolor=blue}

\colorlet{red}{black}

\title[Constraints on B0 from the TNG simulation]{Revision of Faraday rotation measure constraints on the primordial magnetic field using the IllustrisTNG simulation}

\author[A. Arámburo-García et al.]{Andrés Arámburo-García$^{1}$\thanks{aramburo@lorentz.leidenuniv.nl},
Kyrylo Bondarenko$^{2,3,4}$\thanks{kyrylo.bondarenko@sissa.it},
Alexey Boyarsky$^{1}$\thanks{boyarsky@lorentz.leidenuniv.nl},
Andrii Neronov$^{5,6}$\thanks{andrii.neronov@apc.in2p3.fr},
\newauthor
Anna Scaife$^{7}$\thanks{anna.scaife@manchester.ac.uk} and Anastasia Sokolenko$^{8,9}$\thanks{ sokolenko@kicp.uchicago.edu}
\\
$^{1}$Institute Lorentz, Leiden University, Niels Bohrweg 2, Leiden, NL-2333 CA, the Netherlands\\
$^{2}$IFPU, Institute for Fundamental Physics of the Universe, via Beirut 2, I-34014 Trieste, Italy\\
$^{3}$SISSA, via Bonomea 265, I-34132 Trieste, Italy\\
$^{4}$INFN, Sezione di Trieste, SISSA, Via Bonomea 265, 34136, Trieste, Italy\\
$^{5}$Universit\'e  de  Paris  Cite,  CNRS,  Astroparticule  et  Cosmologie, F-75013  Paris,  France\\
$^{6}$Laboratory  of  Astrophysics,  Ecole  Polytechnique  Federale  de  Lausanne,  CH-1015,  Lausanne,  Switzerland\\
$^{7}$Department of Physics \& Astronomy, University of Manchester, UK\\
$^{8}$Theoretical Astrophysics Department, Fermi National Accelerator Laboratory, Batavia, Illinois, 60510, USA\\
$^{9}$Kavli Institute for Cosmological Physics, The University of Chicago, Chicago, IL 60637, USA
}

\date{Accepted XXX. Received YYY; in original form ZZZ}

\pubyear{2022}

\begin{document}

\label{firstpage}
\pagerange{\pageref{firstpage}--\pageref{lastpage}}
\maketitle


\begin{abstract}
\textcolor{red}{Previously derived Faraday rotation constraints on the volume-filling intergalactic magnetic field (IGMF)} have used analytic models that made a range of simplifying assumptions about magnetic field evolution in the intergalactic medium and did not consider the effect of baryonic feedback on large-scale structures.
\textcolor{red}{In this work we revise existing Faraday rotation constraints on the IGMF using a numerical model of the intergalactic medium from the IllustrisTNG cosmological simulation that includes a sophisticated model of the baryonic feedback.}
We use the IllustrisTNG model to calculate the rotation measure and compare the resulting mean and median of the absolute value of the rotation measure with data from the NRAO VLA Sky Survey (NVSS). The numerical model of the intergalactic medium includes a full magneto-hydrodynamic model of the compressed primordial magnetic field as well as a model of the regions where the magnetic field is not primordial, but is rather produced by the process of baryonic feedback. Separating these two types of regions, we are able to assess the influence of the primordial magnetic field on the Faraday rotation signal. 
\textcolor{red}{We find that by correcting for regions of compressed primordial field and accounting for the fact that part of the intergalactic medium is occupied by magnetic fields spread by  baryonic feedback processes rather than by the primordial field relaxes the Faraday rotation bound by a factor of $\simeq 3$. This results in $B_0< 1.8\times10^{-9}$~G for large correlation length IGMFs.}
\end{abstract}


\section{Introduction}

Magnetic fields generated by various types of dynamos in galaxies and galaxy clusters may originate from yet uncertain weak "seed" magnetic fields that are relic of the Early Universe \citep{1994RPPh...57..325K,2001PhR...348..163G,Durrer:2013pga,2021RPPh...84g4901V}. In this case, the relic magnetic fields not affected by dynamos may still reside in the intergalactic medium outside galaxies and galaxy clusters. Detection of such fields may shed light on some of the unresolved problems of cosmology, such as the baryon asymmetry of the Universe. 
The mere presence of the baryon asymmetry suggests that some non-equilibrium process happened in the Early Universe to fulfill one of the Sakharov conditions. Departure from thermal equilibrium is also a necessary condition for the generation of the magnetic field that may thus encode information on the details of a process responsible for the formation of matter-antimatter asymmetry within the first microseconds of the existence of the Universe.
Reliable detection of relic magnetic fields in the present-day Universe would constitute a new pillar of cosmology -- a new relic observable, more ancient than those we can measure so far.

The only place in the Universe where primordial magnetic fields untouched by contamination from galactic feedback may be found in the intergalactic medium (IGM). Magnetic field strength in the IGM is currently constrained from below by gamma-ray observations \citep{Neronov:1900zz} and from above by the Faraday rotation \citep{Blasi:1999hu,Pshirkov:2015tua} as well as by CMB~\citep{Durrer:2013pga,Planck:2015zrl} and ultra-high-energy cosmic ray data \citep{2021arXiv211208202N}. The Faraday rotation constraint is based on observations of the rotation of the polarization plane of electromagnetic radiation during its propagation through a medium containing both free charges and a magnetic field. The total rotation angle is proportional to the Rotation Measure (RM), which is an integral along the line of sight of the product of the free charge number density and the magnetic field component parallel to the line of sight. For an extragalactic source, one can typically distinguish three contributions to this integrated quantity: (i) the local medium near the source, (ii) the IGM, and (iii) our Galaxy.  
The Galactic RM often provides a dominant contribution \citep[see e.g.][]{Oppermann:2014cua}. Uncertainties in modeling the Galactic magnetic field induce a large systematic uncertainty to the estimates of the "residual" RM (RRM) signal obtained after subtracting the Galactic RM component from the total RM toward extragalactic sources. 

The constraint on the primordial magnetic field is obtained from comparing the RRM data with model predictions of the rotation measure signal produced by the volume-filling primordial field. These predictions are also uncertain because they have to rely on a model of the distribution of free electrons in the intergalactic medium and on the model of the cosmological magnetic field. Both the free electron cosmological magnetic field is transformed by the Large Scale Structure formation process. An analytical model of the intergalactic medium has been proposed by  \citet{Blasi:1999hu}. This analytical model represents the IGM as a set of regions of fixed size equal to the Jeans length, with electron density $n_e$  in different regions distributed following a log-normal distribution derived from statistics of Ly$\alpha$ forest data. In this model, the magnetic field strength scales as $n_e^{2/3}$, assuming adiabatic contraction respecting magnetic flux conservation. Using this model,  \cite{Blasi:1999hu} have derived a constraint on the IGMF by comparing the redshift dependence of the median RM from their theoretical predictions with the observational RM catalog of \cite{1984ApJ...279...19W}. \citet{Pshirkov:2015tua} have updated the constraint of \citet{Blasi:1999hu} using the same analytical model and a larger RM dataset of  \citet{2012arXiv1209.1438H} that is based on the RM data derived from NRAO VLA Sky Survey, NVSS \citep{Taylor2009}. The upper bound on IGMF derived from the Faraday rotation data depends on the correlation length. \textcolor{red}{On intermediate scales ($0.5-3.0$\,Mpc), IGMF upper limits of order $10-100$\,nG have been found from RM data \citep[e.g.][]{OSullivan:2020pll,Mtchedlidze:2021bfy,Amaral:2021mly,carretti:2022}.} For magnetic fields with larger correlation lengths (nearly homogeneous on scales comparable to the Hubble radius), the limit derived by \citet{Pshirkov:2015tua} is $B<0.65$\,nG. 

A more advanced approach for modeling the magnetic field in the intergalactic medium is implemented in cosmological simulations based on magneto-hydrodynamics (MHD), like IllustrisTNG simulation \citep{marinacci18,Nelson2019ComAC...6....2N}. In such simulations, the flux-conserving scaling of magnetic field $B\propto n_e^{2/3}$ is found automatically in the regions undergoing adiabatic contraction in the process of galaxy formation. In addition to the adiabatic contraction scaling, cosmological MHD simulations capture other effects that were not considered in the analytical models of  \citet{Blasi:1999hu,Pshirkov:2015tua}, e.g. turbulent dynamo~\citep{2011ApJ...738..134A,Akahori:2014uza}. The magnetic field in the IGM is influenced by feedback processes by galaxies that lead to the formation of magnetized bubbles around galaxies \citep{Bertone:2006mr}. \cite{Garcia:2020kxm} have considered this effect in  IllustrisTNG simulation and found that the magnetized bubbles around galaxies occupy up to $15\%$ of the simulation volume. Preliminary estimates of \citet{Garcia:2020kxm} show that the contribution to the RM from these magnetic bubbles found along the lines of sight toward extragalactic sources can be comparable to the  Galactic RM. The presence of the magnetized bubbles reduces the space occupied by the primordial magnetic field and makes its detection through the Faraday rotation effect more challenging.

In this work, we revise the previously derived upper bound on the volume filling IGMF using the IllustrisTNG numerical model of the IGM instead of the analytical model of \citet{Blasi:1999hu}. 
Considering the same dataset as \cite{Pshirkov:2015tua} (\citet{2012arXiv1209.1438H}, with an updated Galactic RM model of \cite{Hutschenreuter2020}), we find that the Faraday rotation bound on IGMF is relaxed by a factor of $\simeq 3$ after the account of the more precise model of the compressed primordial field and the presence of magnetized bubbles produced by the baryonic feedback.

The structure of this paper is as follows: in Section~\ref{sec:RM-divergence} we describe how the mean rotation measure depends on the magnetic field in the regions occupied by the cosmological field and describe how the electron number density distribution influences this quantity. This discussion demonstrates the need to use cosmological simulations to predict the RMs. In Section~\ref{sec:bubbles} we describe the IllustrisTNG simulations and the properties of magnetic bubbles where the galactic feedback results in amplification due to dynamo effects. 
In Section~\ref{sec:bubble-prim-division} we discuss the separation of the volume-filling component of the IGM from that of the magnetic bubbles and extract a prediction for the RM from the volume-filling magnetic field component. In Section~\ref{sec:constraints} we test the model predictions against the observational data to derive constraints on the magnetic field strength. In the last Section~\ref{sec:discussion} we describe the implications of these results, and we draw our conclusions. 

Cosmological parameters from \cite{Plank2016A&A...594A..13P} are assumed throughout this work.

\section{Mean rotation measure for the primordial magnetic field}
\label{sec:RM-divergence}

The average value of the volume-filling magnetic field changes with the scale factor of the Universe, $a$, as $1/a^2$. In what follows, we refer to the "comoving" magnetic field strength in which the scale factor dependence is factored out. 
The local value of the field can be significantly different from the average. 
The field is transformed by the process of structure formation. Its strength is enhanced by the adiabatic contraction proportional to $n_{\rm e}^{2/3}$. 
The compressed field may be further transformed by various types of dynamos that may be active in the course of the structure formation process \citep{Dolag:2004kp,Vazza:2014jga,Vazza:2021vwy}.

The adiabatic contraction influences the RM, which  is the integral over the line of sight
\begin{equation}
    \text{RM} = \frac{e^3}{2\pi m_{\rm e}^2 \textcolor{red}{c^4}}\int \frac{n_{\rm e} B_{\parallel}}{(1+z)^2} \frac{{\rm d}\ell}{{\rm d}z} {\rm d}z,
    \label{eq:RMeq}
\end{equation}
where $e$ and $m_{\rm e}$ are the electron charge and mass, while $n_{\rm e}$ and $B_{\parallel}$ are the free electron number density and the parallel component of the magnetic field along the line of sight. Adiabatic contraction of regions along the line of sight leads to the increase of the mean RM along the line of sight.

\begin{equation}
    \langle |\text{RM}| \rangle \propto \int\limits_0^{\infty} n_{\rm e}^{5/3} \frac{{\rm d}P}{{\rm d}n_{\rm e}} {\rm d}n_{\rm e},
    \label{eq:avRM}
\end{equation}
where ${\rm d}P/{\rm d}n_{\rm e}$ is the probability density function (PDF) of the electron number density in the Universe.

The mean RM may be significantly influenced by the high-density tail of the PDF if it does not decrease fast enough at large $n_e$. 

\begin{figure*}
    \centering
    \includegraphics[width=0.48\textwidth]{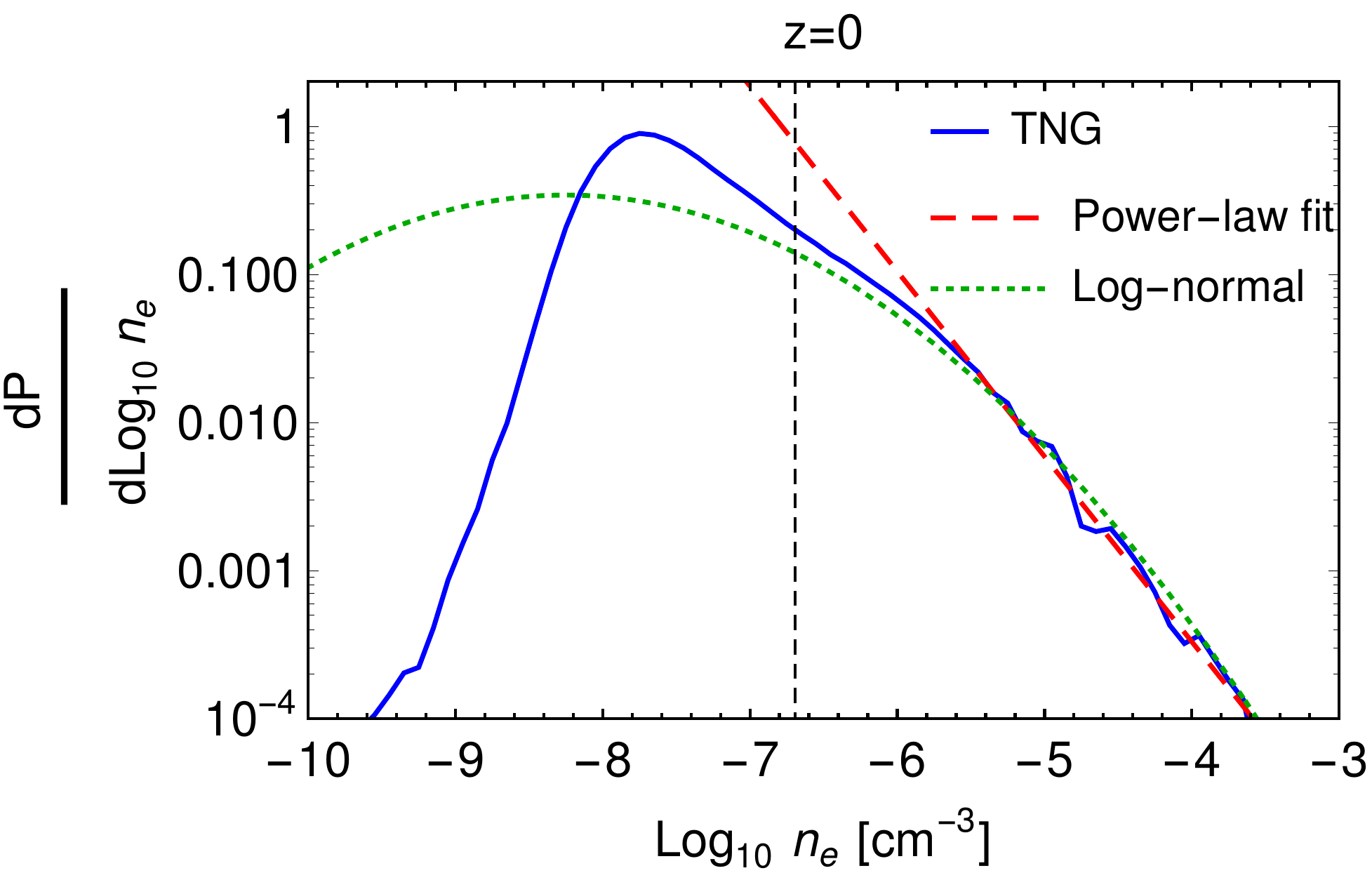}~\includegraphics[width=0.48\textwidth]{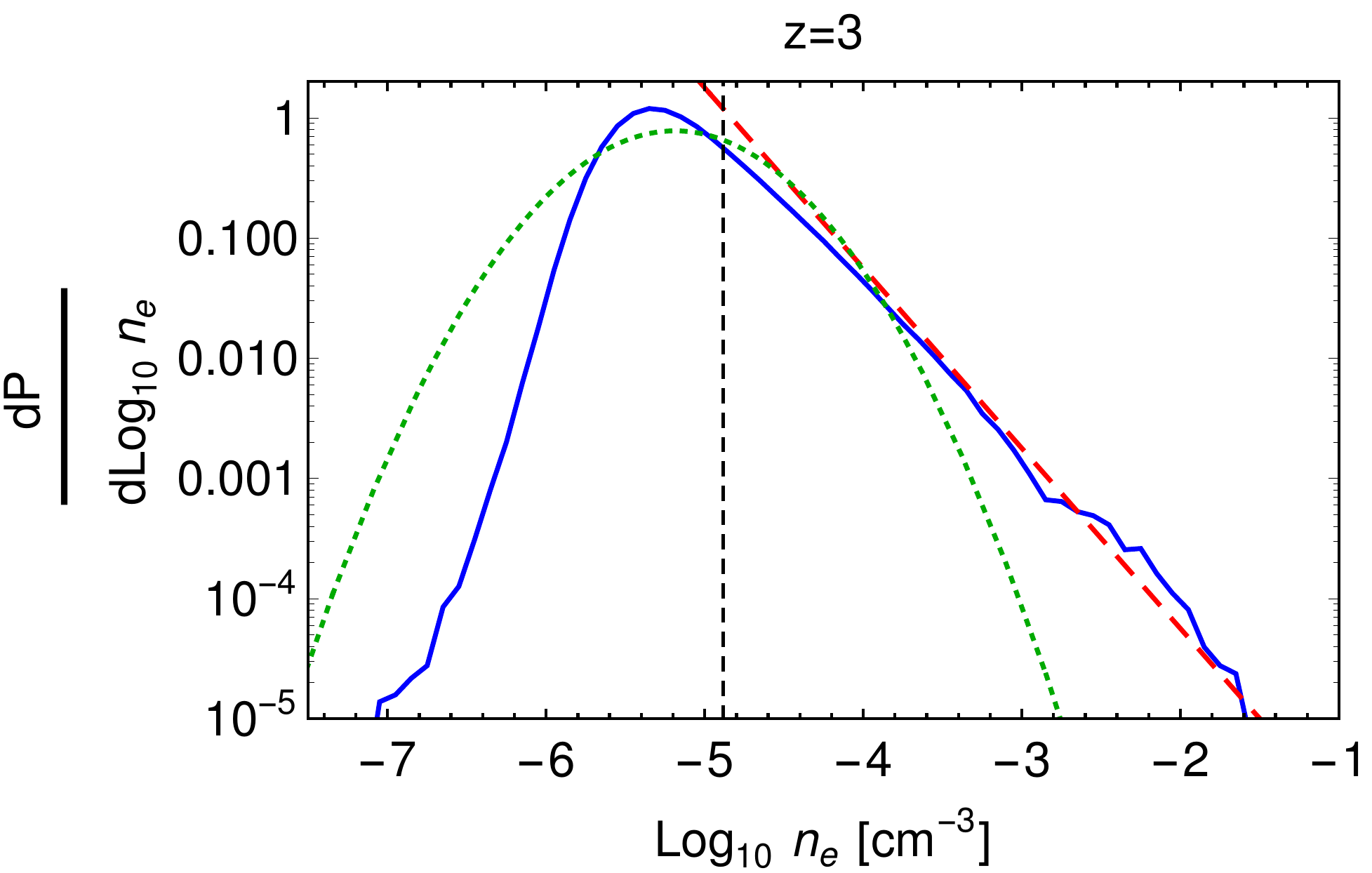}
    \caption[]{Probability distribution for electron number density at $z=0$ (left panel) and $z=3$ (right panel) calculated using random lines of sight in the IllustrisTNG simulation with voxel size $(20\text{ kpc})^3$ (blue line). The black dashed line shows the average electron number density, the red dashed line is the best power-law fit to the high density tail of the distribution, with the relationships $dP/dn_{\rm e} \propto n_{\rm e}^{-2.25}$ at $z=0$ and $dP/dn_{\rm e} \propto n_{\rm e}^{-2.5}$ at $z=3$. The green dotted line shows the log-normal model for electron number density used in~\cite{Pshirkov:2015tua}.}
    \label{fig:ne_pdfs}
\end{figure*}

Figure~\ref{fig:ne_pdfs} we show the probability distribution for the electron number density from the IllustrisTNG simulation (see detailed description of the simulation in the next section) at redshifts $z=0$ and $z=3$, compared to the log-normal analytical model used by  \citet{Blasi:1999hu,Pshirkov:2015tua}.  We see that the high-density tail of the distribution in the simulation is well-described by a log-normal model at $z=0$, but at $z=3$ there are significant differences. The high-density tail can be well described by a power-law ${\rm d}P/{\rm d}n_{\rm e} \propto n_{\rm e}^{\alpha}$, with $\alpha = -2.25$ at $z=0$ and $\alpha= - 2.5$ at $z=3$. This implies that the mean RM value given by Equation~\ref{eq:avRM} is divergent at large $n_e$ limit so that the $|\text{RM}|$ integral is dominated by contributions of strongest overdensities along the line-of-sight.
 
The highest overdensities correspond to collapsed structures such as galaxies and galaxy clusters, and these are filled with strong magnetic fields generated by dynamos, which have "forgotten" the primordial magnetic field configuration. Moreover, as we will see in the next Section, around these collapsed structures, the galactic outflows form areas of over-magnetized media: the magnetic bubbles. A reliable calculation of the integral of Eq. (\ref{eq:RMeq}) has to be based on a model that correctly describes the high-density tail of the $n_e$ distribution.  
In what follows, we use the numerical model of magnetized IGM from the IllustrisTNG simulation to estimate the extragalactic contribution to the RM integral.

\section{Description and processing of simulation data}
\label{sec:bubbles}

\subsection{IllustrisTNG simulations}

The IllustrisTNG project is a number of gravo-magnetohydrodynamic simulations developed by the Illustris project~\citep{nelson18,springel18,pillepich18,naiman18,marinacci18}. It is based on the moving-mesh \textsc{Arepo} code \citep{springel2010MNRAS.401..791S} that solves the system of equations for self-gravity and ideal magnetohydrodynamics~\citep{2011MNRAS.418.1392P,2013MNRAS.432..176P}. 
We use the high-resolution TNG100-1 cosmological simulation \citep[hereinafter TNG100; ][]{Nelson2019ComAC...6....2N}, that has a box size $\sim (110~\text{cMpc})^3$ and contains $1820^3$ dark matter particles and an equal number of initial gas cells with masses of $m_{\text{DM}} = 7.5 \times 10^6~{\rm M}_{\odot}$ and $m_{\text{bar}} = 1.4 \times 10^6~{\rm M}_{\odot}$. 

The initial seed magnetic field in this simulation was chosen as a homogeneous magnetic field with strength $B_0 = 10^{-14}$\,cG (comoving Gauss), directed along the $z$-axis in the simulation box. \textcolor{red}{Such magnetic field configuration with superhorizon correlation length could be generated by some mechanism of production primordial magnetic field during the inflationary phase, see e.g.~\cite{Durrer:2013pga,Subramanian:2015lua}.} This seed magnetic field experiences adiabatic contraction during the structure formation process and is strongly amplified by small-scale dynamos in collapsed structures. 

The TNG simulations include comprehensive galaxy and supermassive black hole (SMBH) formation and feedback models~\citep{Weinberger2017MNRAS.465.3291W,Pillepich2018MNRAS.473.4077P}.
Specifically, when the virial mass of a dark matter halo exceeds $\sim 7 \times 10^{10}$\,M$_{\odot}$, an SMBH with an initial mass of $\sim 10^{6}$\,M$_{\odot}$ is placed in the gravitational potential minimum of the halo.
These black holes grow via smooth gas accretion according to the Bondi-Hoyle-Lyttleton model~\citep{2017MNRAS.465.3291W} or via binary mergers with each other. SMBH growth rate depends on black hole mass, local gas density, and the relative velocity between the black hole and surrounding gas. 

There are two regimes of SMBH activity, depending on accretion rate~\citep{Weinberger2017MNRAS.465.3291W}. 
In the high-accretion rate (accretion rate above $\sim 10\%$ of the Eddington limit), energy is continuously injected into the gas around an SMBH, which causes thermal heating of the gas.
In the low-accretion rate state, energy is deposited periodically in the form of the kinetic energy of the surrounding gas once enough energy accumulates via accretion (see \cite{Weinberger2017MNRAS.465.3291W} for additional details). In this second case, each injection event creates a randomly oriented high-velocity kinetic wind, and if averaged over large time intervals, energy injection in this mode becomes isotropic.
Perhaps counter-intuitively, the low accretion rate feedback mode produces the most powerful outflows in the TNG model~\citep{2019MNRAS.490.3234N}. 
A large fraction of the injected kinetic energy in this mode thermalizes via shocks in the surrounding gas, thereby providing a distributed heating channel.

\subsection{Magnetic bubbles in IllustrisTNG simulations}
\label{sec:bubbles-small-B0}

According to the TNG simulations, magnetized bubbles of chemically enriched material emerge at $z \lesssim 2$ as a result of active baryonic feedback from galaxies~\citep{Garcia:2020kxm,Garcia:2021cgu,Bondarenko:2021fnn}. These bubbles are extended regions that can have sizes of order tens of Mpc with magnetic fields over $10^{-12}$~cG, which is orders of magnitude larger than the initial seed field in the TNG simulation.
In~\cite{Garcia:2020kxm} it was shown that at $z=0$ such bubbles occupy  $12-14\%$ of the cosmological volume of the simulation for magnetic fields stronger than $10^{-12}$~G and more than $3\%$ for $B>10^{-9}$~G. 
Comparing simulations where the initial magnetic field varied by several orders of magnitude (but which were otherwise identical), \cite{Garcia:2020kxm,Garcia:2021cgu} showed that the magnetic fields in these bubbles are to a large extent independent of the strength of the initial seed field. 
The strong magnetic fields in these bubbles also "forget" the direction of the initial seed field. In contrast, the volume-filling compressed primordial field in the regions outside the bubbles prefers the initial direction \citep{Bondarenko:2021fnn}. 
These magnetized bubbles are mainly caused by the feedback from active galactic nuclei (AGN), with a smaller contribution arising from supernovae \citep[SNe;][]{Garcia:2020kxm}.

\begin{figure}
    \centering
    \includegraphics[width=0.48\textwidth]{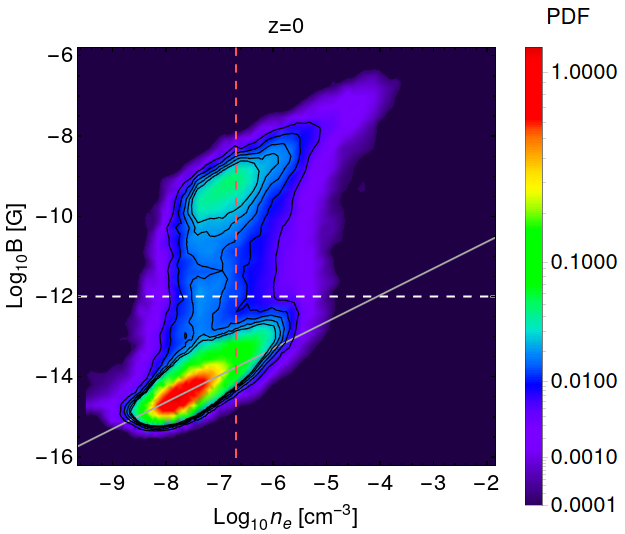}
    \caption[]{The volume-weighted distribution of the magnetic field strength and electron number density in the TNG100 simulation at $z=0$. The dashed white line corresponds to the comoving magnetic field value $10^{-12}$~cG. The red dashed line represents the average electron number density at a given redshift. The gray dashed line shows a power-law $B\propto n_{\rm e}^{2/3}$ that represents adiabatic evolution. The figure is adapted from~\cite{Garcia:2020kxm}.}
    \label{fig:ne_B}
\end{figure}

Fig. \ref{fig:ne_B} shows the distribution of $n_e, B$ values across the simulation volume. One can clearly identify the voxels occupied by the compressed primordial field by the $B\propto n_e^{2/3}$ scaling passing through the reference $B_0=10^{-14}$~cG initial homogeneous field that was taken as the initial condition in the simulation. The voxels in which the field does not follow this scaling are situated close to galaxies and are all affected by the feedback process. In what follows, we "flag" these voxels as belonging to the magnetized bubbles and hence not carrying information about the primordial field. In practice, we flag the voxels as belonging to the magnetized bubbles if the magnetic field strength there is higher than  $10^{-12}$~cG. This effectively leaves all the voxels with the compressed primordial field "un-flagged", see Fig~\ref{fig:ne_B}. The $B>10^{-12}$~cG cut would have to be shifted to a stronger threshold field value if the initial magnetic field in the simulations were stronger. However, only a small number of voxels would be affected by the cut change. The bulk of the voxels affected by the feedback belongs to a higher $B\propto n_e^{2/3}$ branch in Fig.  \ref{fig:ne_B} that corresponds to the magnetic flux conservation in the galactic outflows that produced the magnetized bubbles. Properties of the field in the voxels from this branch are independent of the configuration of the seed magnetic field. They are determined by the physics of the galactic feedback~\citep{Garcia:2020kxm,Garcia:2021cgu}. The contribution of these voxels in the RM integral cannot be considered in the estimate of the RM produced by the primordial field. All the high-density tail of the $n_e$ distribution discussed in Section \ref{sec:RM-divergence} belongs to this high-density tail. Excluding these voxels from the calculation of the RM integral due to the primordial field thus solves the problem of divergence of the integral in the high $n_e$ limit.

\begin{figure}
    \centering
    \includegraphics[width=0.48\textwidth]{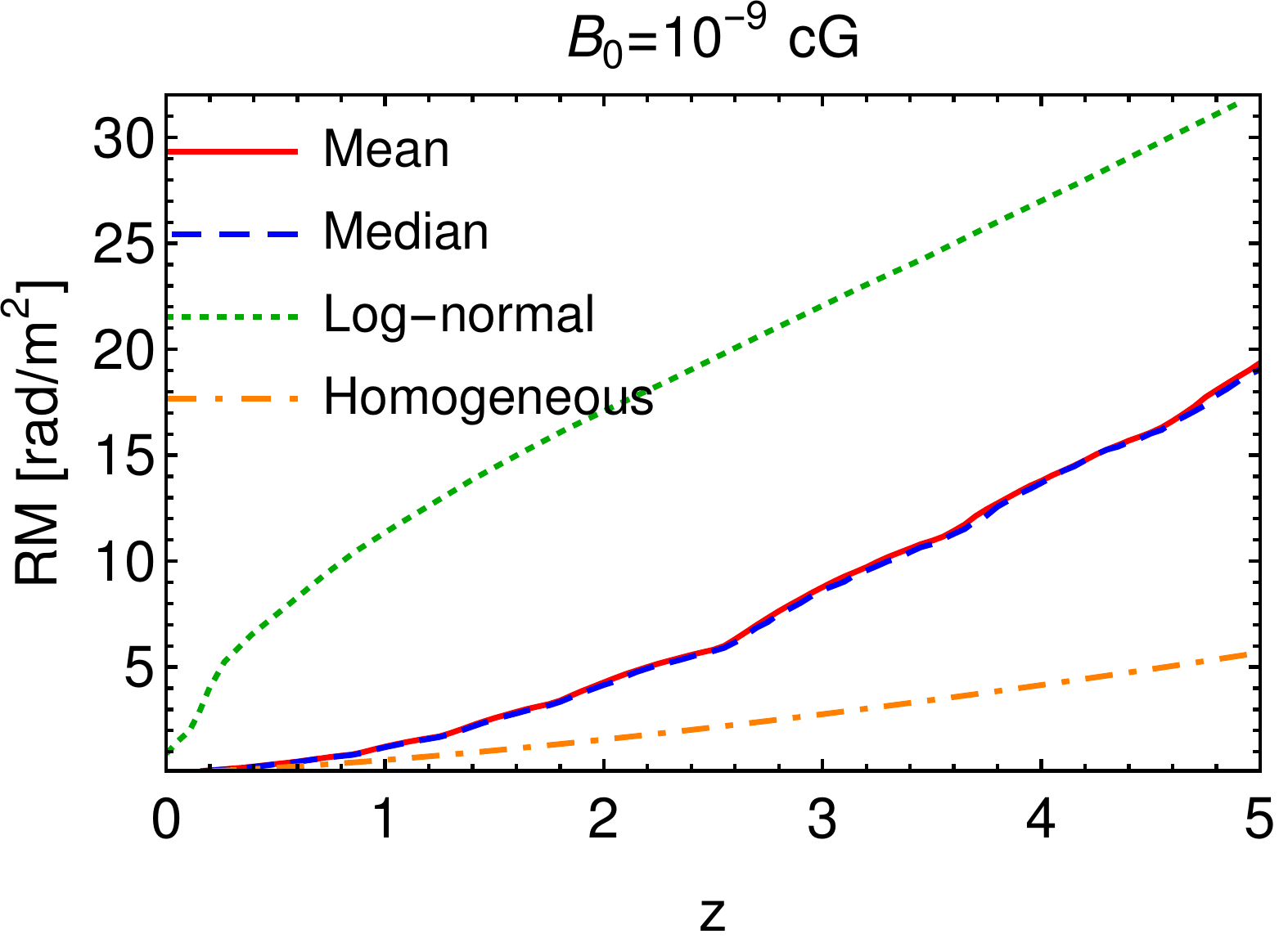}
    \caption[]{Prediction for the mean (blue continuous line) and median value (red dashed line) of $|\text{RM}|$ from the IllustrisTNG simulation as a function of redshift for a homogeneous primordial magnetic field with $B_0 = 10^{-9}$~cG. For comparison we also show the prediction from~\cite{Pshirkov:2015tua} for the mean $|\text{RM}|$ (green dotted line), where the RM was estimated based on an analytic log-normal distribution for electron number density and prediction for the mean $|\text{RM}|$ for the homogeneous Universe (orange dot-dashed line).}
    \label{fig:primordial-prediction}
\end{figure}

\section{Rotation measure from the primordial field}
\label{sec:bubble-prim-division}

The lower $B\propto n_e^{2/3}$ branch in Fig. \ref{fig:ne_B} corresponds to the voxels occupied by the compressed primordial field. To estimate the contribution of the compressed primordial magnetic field to the RM integral, we consider only these voxels. Even though the original IllustrisTNG simulation was done with a fixed initial field strength $B_0=10^{-14}$cG, the scaling of the RM integral with $B_0$ can be readily deduced from the fact that the field behavior is well-described by the model of adiabatic contraction. 
To calculate the RM due to the primordial field as a function of $B_0$, we simply rescale the magnitude of the magnetic field by a factor $B_0/10^{-14}$~cG, retaining the direction of the field. We then use the resulting configuration of the magnetic field to predict the contribution of the primordial magnetic field.
Using this approach, we generate 1000 continuous random lines of sight from $z=0$ to $z=5$ for the electron number density and magnetic field, as described in Appendix~\ref{sec:RM-sim}, and use these data to represent the primordial regions.

In Figure~\ref{fig:primordial-prediction} we show predictions for the mean and median absolute RM value, $|\text{RM}|$, for the primordial magnetic field using $B_0 = 10^{-9}$~cG. It is interesting to note that both contributions to the average and median RM from the primordial magnetic field are well approximated by a simple power law, RM$(z) = a\,z^{b}$, with $a = 1.37\text{ rad}/\text{m}^2$ and $b = 1.65$. For comparison, we also show the analytical model results from~\cite{Pshirkov:2015tua}. One can see that the analytical model shows a notably different redshift dependence compared to the numerically calculated primordial RM component. The rapid growth of the RM at $z<1$ in the analytical model is dominated by the contribution of the high-density tail of the $n_e$ distribution. These regions are mostly polluted by the baryonic feedback fields and are considered as a separate contribution to the RM that is "background" for the measurement of the primordial RM component (in the same sense as the Galactic RM measure is a background on top of which the primordial field RM is searched). The high-density tail is absent at high-redshifts in the analytical model (see Fig. \ref{fig:ne_pdfs}). There is no such sharp difference between the high-density tails of $n_e$ distributions at low and high redshifts in the numerical model of IllustrisTNG. The main result of the account of the density increase and the $B\propto n_e^{2/3}$ scaling of the magnetic field in the numerical model is an overall increase of the RM with respect to the estimate for a perfectly homogeneous Universe (dashed-dotted line in Fig. \ref{fig:primordial-prediction}), but there is no qualitative change of the redshift dependence of RM.

\section{Constraints on the primordial magnetic field}
\label{sec:constraints}

In this Section, we compare the IllustrisTNG model estimate of the RM due to the primordial magnetic field found above with the observational data from NRAO VLA Sky Survey~\citep{Condon1998,Taylor2009}, considering two statistical estimators (mean and median values).

\subsection{Observational data}
\label{sec:extragalactic_rotation_measure}

We use ~\cite{2012arXiv1209.1438H} catalog of Faraday rotation measures and redshifts for 4003 extragalactic radio sources derived from the NRAO VLA Sky Survey~\citep{Condon1998,Taylor2009} RM catalog. Following  \cite{2012arXiv1209.1438H} we remove objects close to the Galactic plane ($\ell < 20^\circ$) and consider 3650 sources at higher Galactic latitude. 
It is important to notice that the data for rotation measure in the catalog were measured at two close frequencies and, therefore, may be subject to a wrapping uncertainty~\citep{Taylor2009}. Practically speaking, this means that from the data on polarization angle the experiment cannot distinguish RMs that differ by integer multiples of $\delta\text{RM} = 652.9 \text{ rad}/\text{m}^2$. In~\cite{Taylor2009} additional information on the depolarization was used to resolve this uncertainty. However, the depolarization method does not work for large $|\text{RM}|>520\text{ rad}/\text{m}^2$ because the quantity estimated using depolarization is degenerate for large and small RMs. In such cases, the RM with a smaller absolute value is taken.

To find the extragalactic contribution to the rotation measure, we first subtract the Galactic RM (GRM) in order to obtain the residual rotation measure (RRM). To do this, we use the recent model of the Galactic RM from~\cite{Hutschenreuter2020}, taking the value of the closest voxel in that model to each source as the GRM.

\begin{table*}
\centering
\begin{tabular}{|l|r|r|r|r|r|r|r|r|r|}
\hline
Bin number & 1 & 2 & 3 & 4 & 5 & 6 & 7 & 8 & 9 \\
\hline
Upper bound for $z$ & $0.13$ & $0.40$  & $0.71$  & $1.02$  & $1.36$  & $1.81$ & $2.4$ & $3$ & $5$   \\ \hline
Object number & $564$  & $781$   & $500$   & $441$   & $426$   & $437$   & $322$ & $129$ & $49$   \\ \hline
$\langle |\text{RRM}|\rangle$, rad$/$m$^2$ &
$13.48$ & $12.60$ & $14.23$ & $14.15$ & $15.53$ & $16.39$ & $15.22$ & $16.98$ & $18.08$ \\ \hline
$\Delta\langle |\text{RRM}|\rangle$, rad$/$m$^2$ & $0.62$ & $0.56$ & $0.85$ & $0.75$ & $0.96$ & $1.39$ & $1.69$ & $1.87$& $3.45$ \\ \hline
$\text{Med}\,|\text{RRM}|$, rad$/$m$^2$ &
$8.74$ & $8.77$ & $8.72$ & $9.07$ & $9.41$ & $9.38$ & $8.09$ & $10.77$ &
$10.97$ \\ \hline
$\Delta\text{Med}\,|\text{RRM}|$, rad$/$m$^2$ &
$0.78$ & $0.70$ & $1.07$ & $0.94$ & $1.20$ & $1.74$ & $2.12$& $2.34$ & $4.32$ \\ \hline
\end{tabular}
\caption{Summary of the observational data in bins used in this work. The first row shows the bin number, the second row shows upper bound on each redshift bin (the lower bound of the first bin is $z=0$). In the third row we show the number of observed objects, also we show the mean $|\text{RM}|$ and $|\text{RRM}|$ and their errors in each bin.}
\label{tab:zbins-new}
\end{table*}

\subsection{Constraining procedure}
\label{sec:method}

There are two contributions to the extragalactic RM: (i) the RM of the host galaxy and/or galaxy cluster, and (ii) the contribution to the RM along the line of sight between the source and our Galaxy, which consists of the primordial magnetic field and magnetic bubbles. The latter contribution may also contain RM contributions from galaxies and/or galaxy clusters located between the source and our Galaxy.

We assume that the contributions from host galaxies,  magnetic bubbles, and the primordial magnetic field are not correlated with each other. This means that, on average, every contribution only increases the total extragalactic RM. Consequently, a robust condition is that one individual contribution (primordial field or magnetic bubbles) should not exceed the observed RRM. 

We divide the source set into nine redshift bins, each containing an approximately equal number of objects at low redshift ($\sim500$) up to redshift $z\simeq 1$, see Table~\ref{tab:zbins-new}. Higher redshift bins contain lower numbers of objects. 
For each bin, we calculate mean and median $|\text{RRM}|$ values from the observational data and estimate the associated statistical error, see Appendix~\ref{app:errors} for details. These results are given in Table~\ref{tab:zbins-new}.

To properly compare predictions with observations, we need to take into account the wrapping uncertainty introduced in Section~\ref{sec:extragalactic_rotation_measure}, which affects large RM values. To imitate this behavior in the simulation data, we apply a wrapping correction similar to that used for the observational data. The details of this wrapping correction procedure are given in Appendix~\ref{sec:wrapping-correction}.

\subsection{Results}
\label{sec:results}

\begin{figure}
    \centering
    \includegraphics[width=\linewidth]{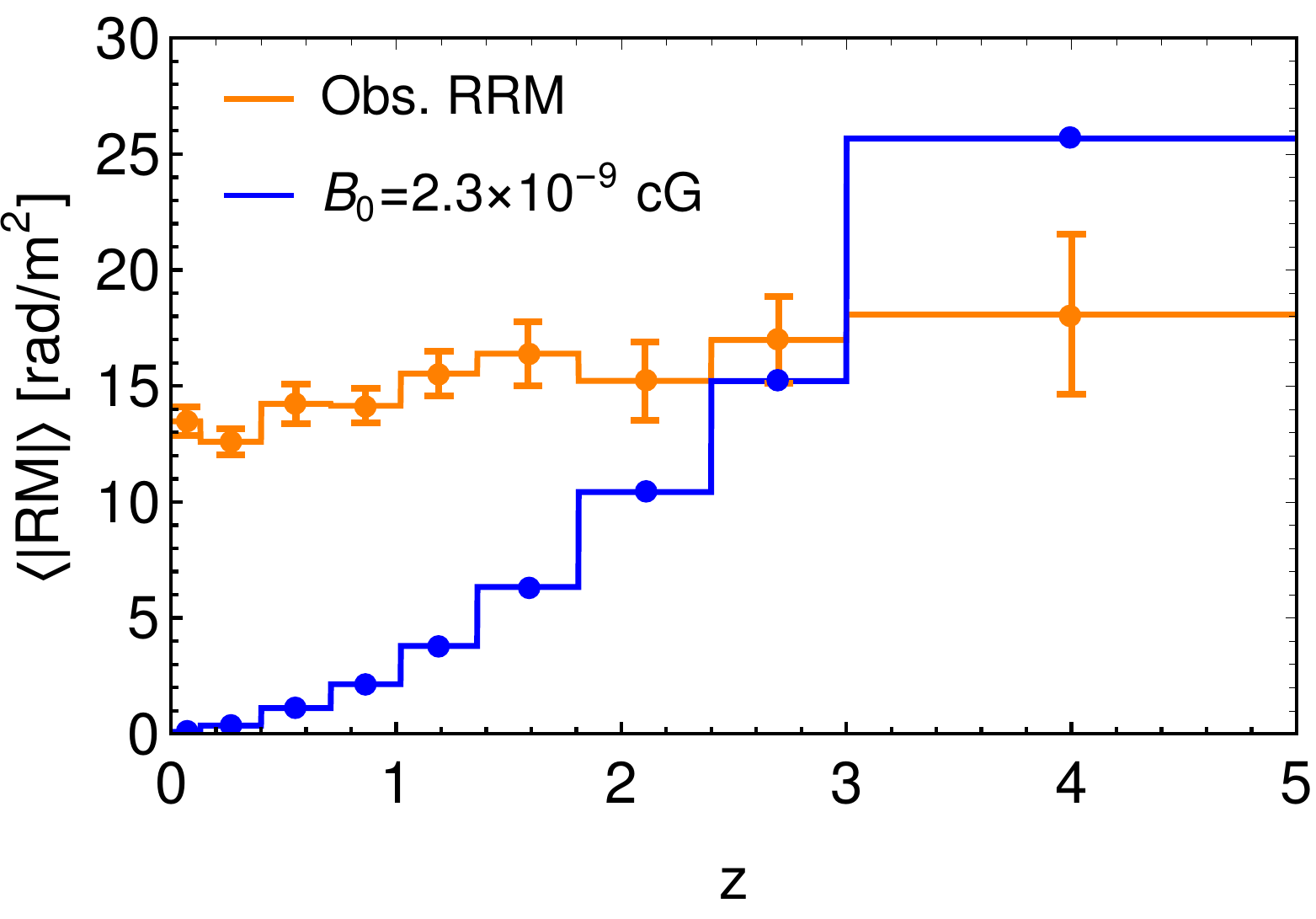}\\
    \includegraphics[width=\linewidth]{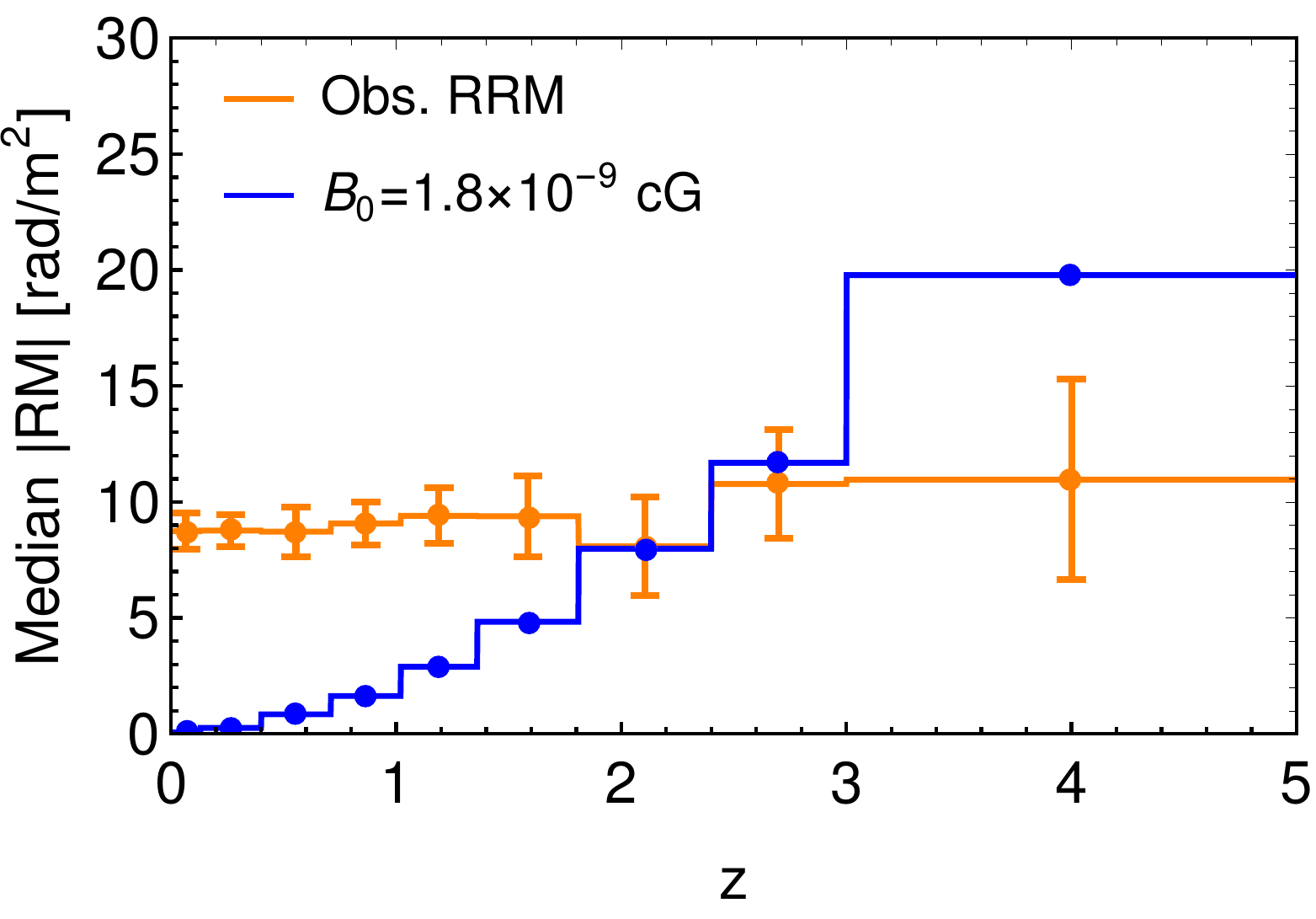}
    \caption{Examples of excluded models for the mean value (upper panel) and for the median value (lower panel) of the primordial magnetic field (blue lines) comparing to the observational data (orange lines).}
    \label{fig:excluded-models}
\end{figure}

Using the predictions for the primordial magnetic field from Section~\ref{sec:bubble-prim-division} we are able to place constraints on $B_0$ in the following way. We calculate a model prediction for the mean and median |RM| in each redshift bin and compare the model predictions to the mean and median of the RM for the NVSS data set. 

Fig. \ref{fig:excluded-models} shows the comparison of the IllustrisTNG model of primordial RM component with the observational data. One can see that the data show only weak, if any, dependence on the redshift, while the model predictions steadily grow and reach the highest value in the highest redshift bin. It is the highest redshift bin that provides the strongest constraint on possible primordial RM contribution to the RRM. Considering this fact, we derive an upper limit on the volume filling primordial magnetic field as the 95\% limit on the mean or median of the absolute value of the RRM in the highest redshift bin. 
These constrains the strength of the homogeneous primordial magnetic field to be $B_0 < 2.3\cdot 10^{-9}$~cG and $B_0 < 1.8\cdot 10^{-9}$~cG, correspondingly. The IllustrisTNG model for this limiting field is shown in Fig.~\ref{fig:excluded-models}.

For completeness, in Appendix~\ref{sec:variance} we also derive constraints on the primordial magnetic field from the RM dispersion, obtaining a much weaker limit.

\section{Discussion and conclusions}
\label{sec:discussion}

Our revised  Faraday rotation limit on the primordial homogeneous magnetic field is by a factor of three weaker than the limit derived by~\cite{Pshirkov:2015tua}. This difference is explained by revising the model of the magnetized intergalactic medium used for the RM calculations. Previous calculations of \citet{Blasi:1999hu,Pshirkov:2015tua}  have used an analytic model that assumed certain distribution of electron density $n_e$ and did not consider the fact that vast regions around galaxies might be affected by the baryonic feedback process. 
We have instead used the IllustrisTNG model of the intergalactic medium, in which we were able to separate regions with volume-filling primordial magnetic fields from magnetized bubbles produced by the galactic feedback on the large-scale structure. 

Using this separation, we were able to examine the predicted contributions to extragalactic Faraday rotation from the volume-filling magnetic field. We have found that the redshift dependence observed for Faraday rotation due to the volume-filling component differs from that of previous works that assumed a log-normal electron density distribution and adopted an assumption that the primordial field is present in regions near galaxies, see Fig. \ref{fig:primordial-prediction}. 

Revision of the model of the IGM has led to the revision of the RM bound on the strength of the primordial magnetic field. This limit is $B_0< 1.8\times10^{-9}$~cG for the large correlation length field comparable to the Hubble scale. This is a robust and conservative bound that takes into account uncertainty of the influence of the baryonic feedback on the intergalactic medium: the regions potentially affected by the feedback are excluded from the calculation of the RM integral due to the primordial field. 

Our revised bound can be compared to the existing constraints on the primordial volume-filling magnetic field.
Direct measurements of the volume-filling IGMF are currently not possible, but various constraints have been made in the existing literature for the strength of this field, $B_0$, as a function of coherence length. A lower limit on $B_0$ has been established observationally using the non-detection of secondary photon cascades from gamma-ray sources \citep{Neronov:1900zz}. The level of the lower bound for large correlation length fields quoted in different references depends on the assumptions about the properties of gamma-ray sources considered in the analysis. A conservative bound is currently at the level $10^{-17}$~cG  \cite{Taylor:2011bn,2018ApJS..237...32A} based on non-detection of delayed emission from hard spectrum gamma-ray blazars. 

The most constraining direct upper limits on the volume-filling primordial field in the present-day Universe currently come from the anisotropy of ultra-high-energy cosmic rays  \citep[UHECRs;][]{Bray:2018ipq,2021arXiv211208202N}, which results in a limit of $B_0 \lesssim 10^{-10}$\,cG on the uniform component of the IGMF. The UHECR limits have their limitations: the limit of  \cite{Bray:2018ipq} is subject to uncertainty in the composition of UHECRs and the significance of the dipole anisotropy as measured using different UHECR datasets, while the limit of  \cite{2021arXiv211208202N} relies on identification of the UHECR source with the Perseus-Pisces supercluster. The RM limit discussed in our paper is weaker than the UHECR limit, but it is free from the assumptions adopted for the derivation of the UHECR bound, and in this sense, it is complementary to the UHECR bound.

CMB power spectrum analysis produces an upper limit of $B_0 \lesssim 10^{-9}$\,cG on a large-scale field, that extends uniformly across coherence lengths from the Hubble radius down to $\sim 1$\,Mpc \citep{Plank2016A&A...594A..13P}. Various scenarios of the primordial magnetic field can affect this CMB limit, causing it to vary by a factor of $\sim5$, with the most constraining being that of a scale-invariant primordial magnetic field. A tighter bound, $B_0<4.7\times 10^{-11}$~cG, can be derived from the CMB data based on non-observation of the magnetic field induced clumping of baryonic fluid during recombination epoch \citep{2019PhRvL.123b1301J}. Contrary to the RM and UHECR bounds, the CMB bounds are "indirect" in the sense that they constrain the field present during the recombination epoch rather than in the present-day intergalactic medium.

The RM bound on $B_0$ can potentially be improved if a reliable model of the magnetization of the IGM by the baryonic feedback process is available. In this case, it will be possible to subtract the RM component due to the magnetized bubbles around galaxies in the RRM estimate (similarly to the Galactic RM component that is subtracted from the overall RM in the calculation of the RRM). A reliable model of the magnetized galactic outflows can emerge with the new RM data of next-generation radio telescopes, such as SKA and its precursors that will be able to measure the extragalactic RM due to the magnetized bubbles and, in this way, test the feedback models, including the IllustrisTNG model considered in our analysis~\citet{Bubblespaper}.

\section*{Acknowledgements}

KB is partly funded by the INFN PD51 INDARK grant.
AB is supported by the European Research Council (ERC) Advanced Grant ``NuBSM'' (694896). AMS gratefully acknowledges support from the UK Alan Turing Institute under grant reference EP/V030302/1. AS is supported by the Kavli Institute for Cosmological Physics at the University of Chicago through an endowment from the Kavli Foundation and its founder Fred Kavli. This work has been supported by the Fermi Research Alliance, LLC under Contract No. DE-AC02-07CH11359 with the U.S. Department of Energy, Office of High Energy Physics.


\section*{Data Availability}

The data underlying this article is available on reasonable request.

\bibliographystyle{mnras}
\bibliography{refs.bib}

\appendix

\section{Generation of continuous lines of sight}
\label{sec:RM-sim}

To calculate the prediction for rotation measure from the IllustrisTNG simulations, we create continuous lines of sight (LOS) in order to recover the electron number density and the magnetic field up to redshift 5. Our method is the same as that used by \cite{Garcia:2020kxm} and employs the following steps:
\begin{enumerate}
    \item Extract 1000 randomly oriented LOS from all available simulation snapshots in the TNG100-1 simulation. Along these LOS, we calculate electron number density and magnetic field values averaged over $(20\text{ kpc})^3$ voxels. 
    If we expect a substructure inside a given voxel, we use a higher resolution to calculate the RM, see Appendix~\ref{app:strong-voxels} for details.
    \item A continuous LOS is produced by stacking together LOS from each snapshot, taking random lines of sight from the snapshots that are the closest by redshift. Then we emulate redshift dependence by scaling the electron number density at each point by $(1+z)^3$ and the magnetic field value by $(1+z)^2$.
\end{enumerate}

\begin{figure}
    \includegraphics[width=0.48\textwidth]{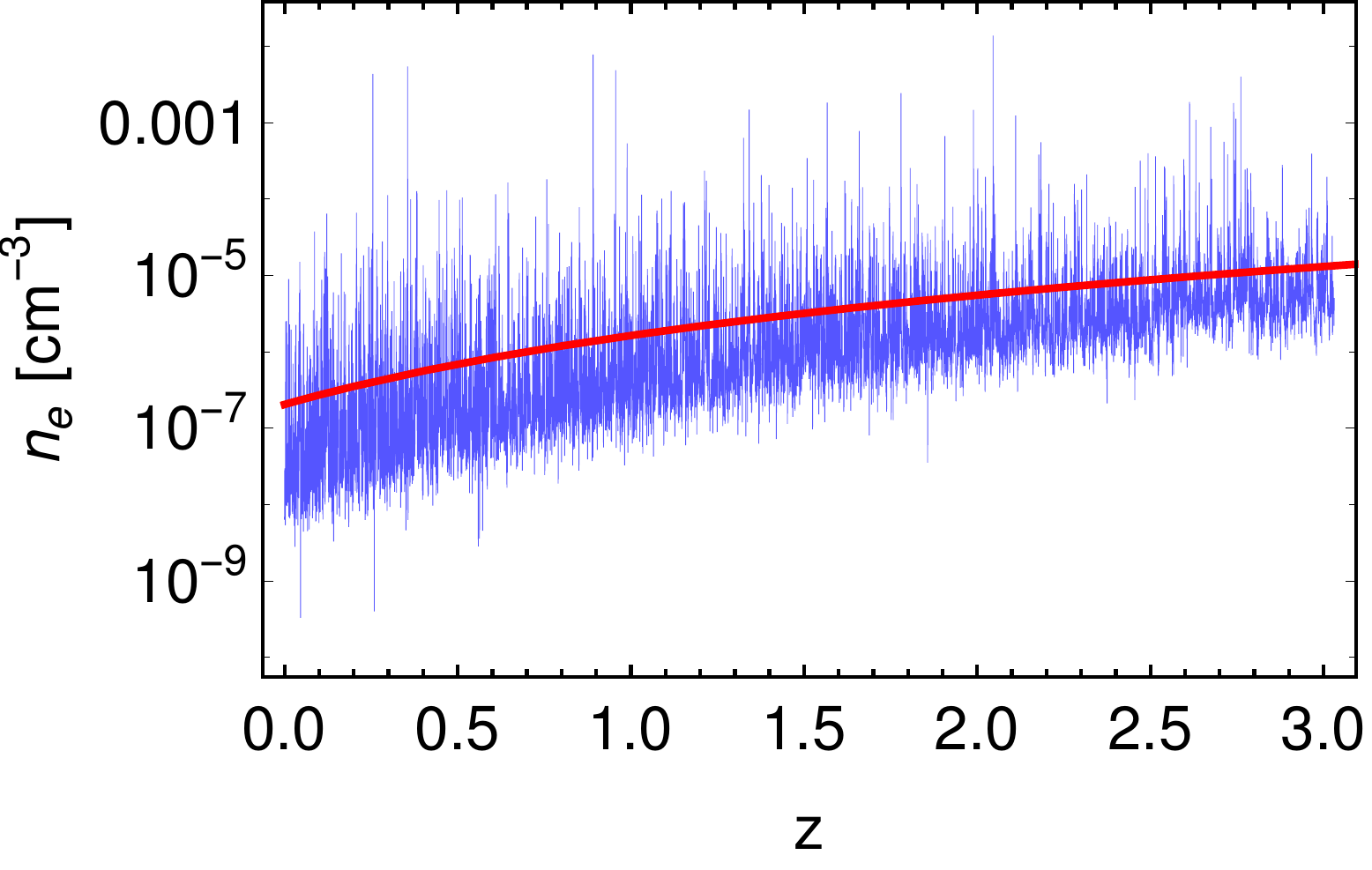}
    \\
    \includegraphics[width=0.48\textwidth]{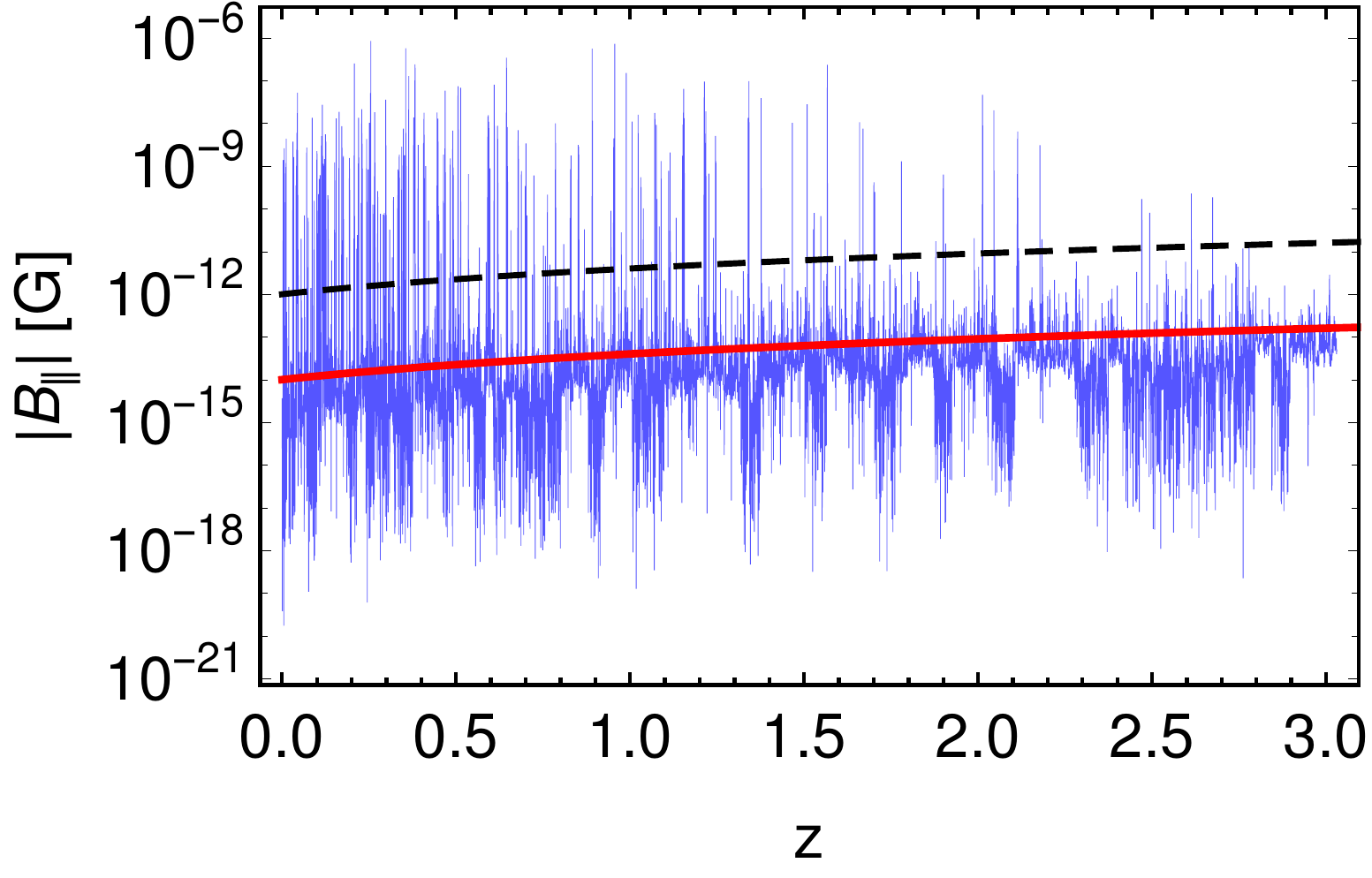}
    \caption{Examples for free electron number density (upper panel) and parallel component of magnetic field (lower panel) along a continuous line of sight generated from data of TNG100-1 simulation. The red line corresponds to the cosmological average value of the electron number density (upper panel) and seed magnetic field value $B_0 = 10^{-14}$~cG used in the simulation (lower panel). The dashed black line in the lower panel shows $B = 10^{-12}$~cG, which we take as a minimal magnetic field value in magnetic bubbles.}
    \label{fig:LOSexamples}
\end{figure}

An example of the continuous LOS produced by our procedure is shown in Figure~\ref{fig:LOSexamples}. \textcolor{red}{Using this procedure we loose large-scale (larger than the simulations box size) correlations. However we believe that should only mildly affect our results.}

\section{Contribution of individual voxels with strong RM}
\label{app:strong-voxels}

\begin{figure}
    \centering
    \includegraphics[width=0.48\textwidth]{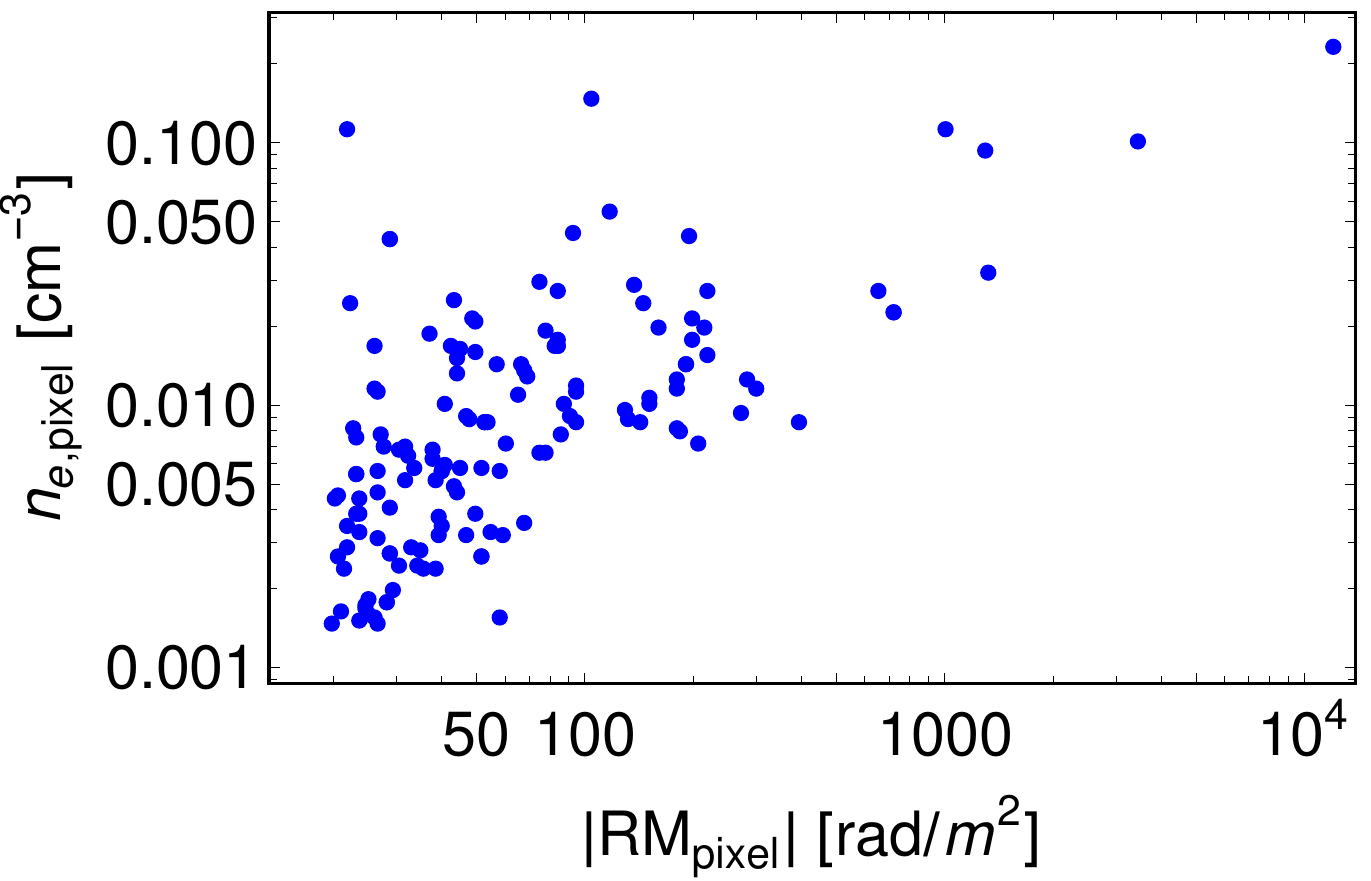}
    \\
    \includegraphics[width=0.48\textwidth]{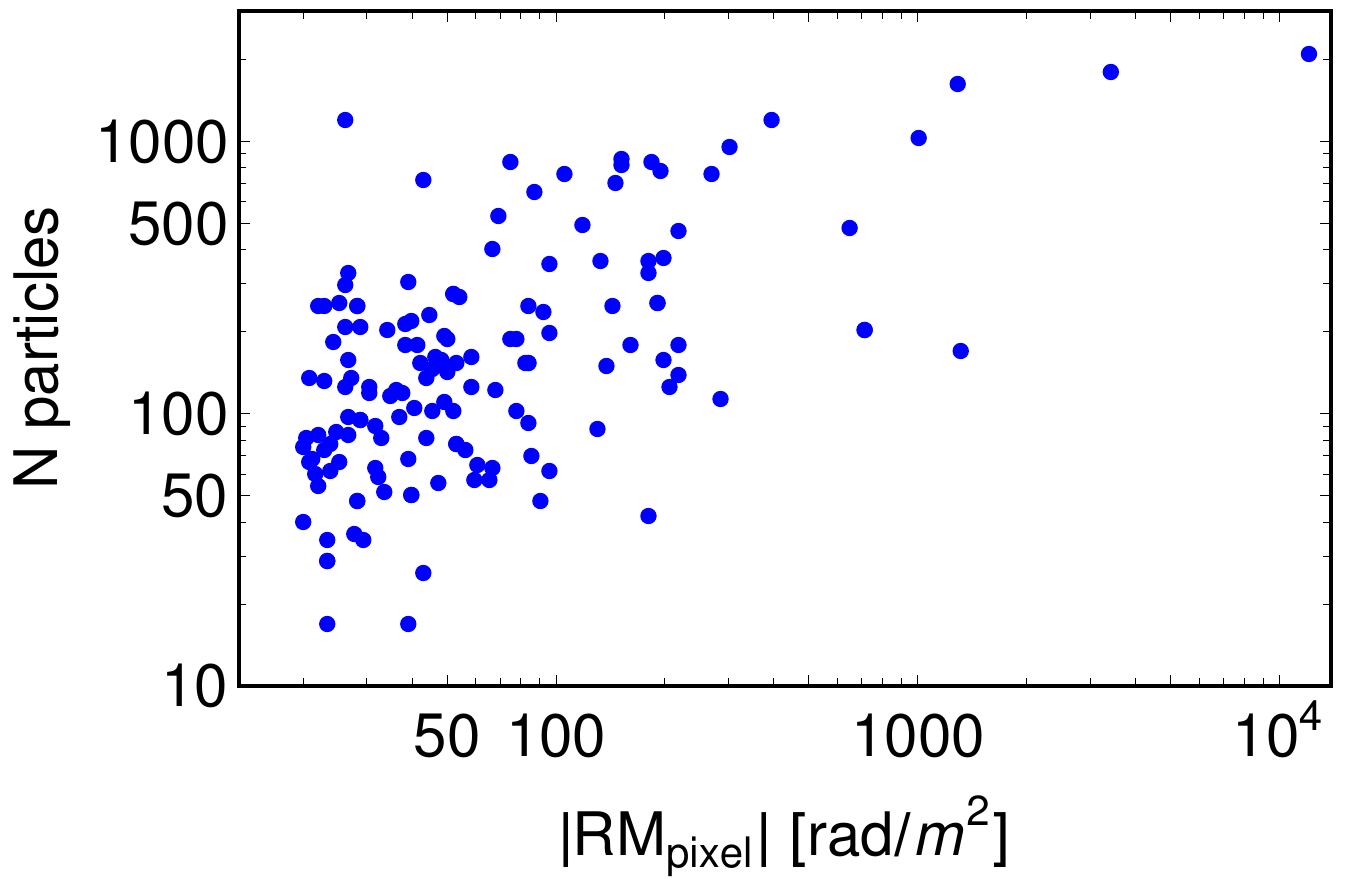}
    \caption{Distribution of electron number density versus RM for voxels with RM$>20\text{ rad}/\text{m}^2$ (upper panel). At lower panel we estimated the number of baryon particles in each of these voxels.}
    \label{fig:ExcludedPixelDistribution}
\end{figure}

In our simulated data we have some voxels in which the RM values $>20~\text{ rad}/\text{m}^2$. Such large RM values could be an artifact of our LOS procedure, as, given our default voxel size ($20$~ckpc), dense substructures such as galaxies are not finely resolved. We have to be cautious since we expect such substructures to provide large RM contributions, and thus even a small fraction of our voxel could overwhelm the rest of the $20$~ckpc volume, even though it is unlikely to be encountered by a realistically random LOS. In the upper panel of Figure~\ref{fig:ExcludedPixelDistribution} we show the distribution of electron number density versus RM for a set of voxels with high RM. We observe that the values of $n_{\rm e}$ for those voxels are typical for galaxies and galaxy clusters. Given these results, it is necessary to add a method to our algorithm that identifies and resolves such voxels at higher resolution. 

As substructures are dense objects, the simulation will assign to their region a large number of gas cells, $N_{\text{part}}$. With this in mind, we can use the average electron number density in the voxel and a baryonic particle mass for the gas cells of $m_b = 1.4\cdot 10^6$\,M$_{\odot}$ to easily estimate $N_{\text{part}}$ within the voxel using
\begin{equation}
    N_{\text{part}} = \frac{(20\text{ kpc})^3}{(1+z)^3} \frac{n_e m_p}{m_b} \approx 1.4\cdot 10^3\ (1+z)^{-3} \left(\frac{n_e}{0.01\text{ cm}^{-3}}\right).
\end{equation}

\begin{figure}
    \centering
    \includegraphics[width=\linewidth]{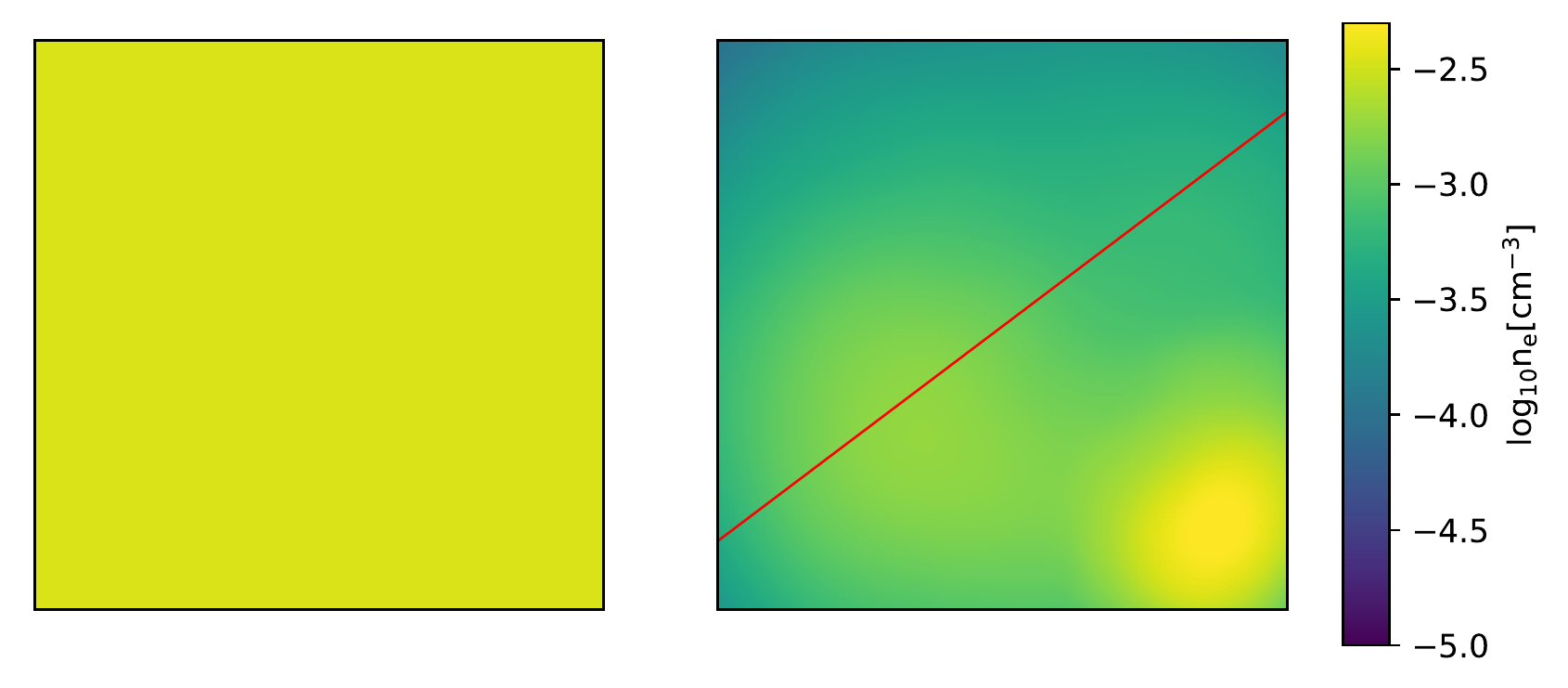}
    \caption{Example region containing gas cells with high $\mathrm{n_e}$. The left panel shows the result of smoothing the region over a single voxel, while in the right panel we use a high resolution grid smoothed over the same region, where only the voxels that lie along the red line will contribute to the calculation of the new RM. The red line preserves the original orientation of the random LOS.}
    \label{fig:highres}
\end{figure}

In the lower panel of Figure~\ref{fig:ExcludedPixelDistribution} we show the estimated $N_{\text{part}}$ for each voxel with large RM. We use this quantity as a threshold in our procedure. A voxel with a $N_{\text{part}} \geq 30$ will be recalculated by creating a new high resolution smoothing grid. From the resulting grid, we will take into account only the contribution of the new sub-voxels that match the orientation of the original LOS, see Figure~\ref{fig:highres}, to which we apply Equation ~\ref{eq:RMeq} and assign that RM result to the original voxel.


\section{Statistical error for mean and median}
\label{app:errors}

For observational data, we calculate the standard error $\Delta x$ for the mean value in each bin as
\begin{equation}
    \Delta \langle x \rangle = \sqrt{\frac{\langle(x_i - \langle x \rangle)^2\rangle}{n}},
\end{equation}
where $x_i$ are $|\text{RRM}|$ values in each bin, $\langle x \rangle$ is their mean value, and $n$ is a number of objects in the bin. In the same notation, the error of the median is estimated as \citep{williams_2001}
\begin{equation}
    \Delta \text{Med}(x) = \sqrt{\frac{\pi}{2} \frac{\langle(x_i - \langle x \rangle)^2\rangle}{n}} \approx 1.253 \cdot \Delta \langle x \rangle.
\end{equation}

\section{Wrapping correction}
\label{sec:wrapping-correction}

As was discussed in Section~\ref{sec:extragalactic_rotation_measure}, the observational data for rotation measure used in this work can be subject to a wrapping uncertainty with step $\delta \text{RM} = 652.9\text{ rad}/\text{m}^2$. In our theoretical prediction, lines of sight sometimes appear with RMs of order $\mathcal{O}(1000)\text{ rad}/\text{m}^2$, so it is important to make a wrapping correction similar to that of the experimental data if we wish to compare them directly. In the case of simulations, we do not have depolarization data available, but based on the description of the procedure, we emulate it in the following way:
\begin{enumerate}
    \item If the absolute value of the RM is smaller than $520\text{ rad}/\text{m}^2$ we do not change it.
    \item If $|\text{RM}|>520\text{ rad}/\text{m}^2$ we take the value $\text{RM} + N \delta\text{RM}$, where $N$ is such integer number (positive or negative) such that the resulting RM has the smallest absolute value.
\end{enumerate}
A few examples:
\begin{itemize}
    \item If RM$=500\text{ rad}/\text{m}^2$ the result is $500\text{ rad}/\text{m}^2$;
    \item If RM$=600\text{ rad}/\text{m}^2$ the result is $-52.9\text{ rad}/\text{m}^2$;
    \item If RM$=700\text{ rad}/\text{m}^2$ the result is $47.1\text{ rad}/\text{m}^2$;
    \item If RM$=1000\text{ rad}/\text{m}^2$ the result is $-305.8\text{ rad}/\text{m}^2$.
\end{itemize}

\section{Constraints from RM dispersion}
\label{sec:variance}

\begin{figure}
    \centering
    \includegraphics[width=\linewidth]{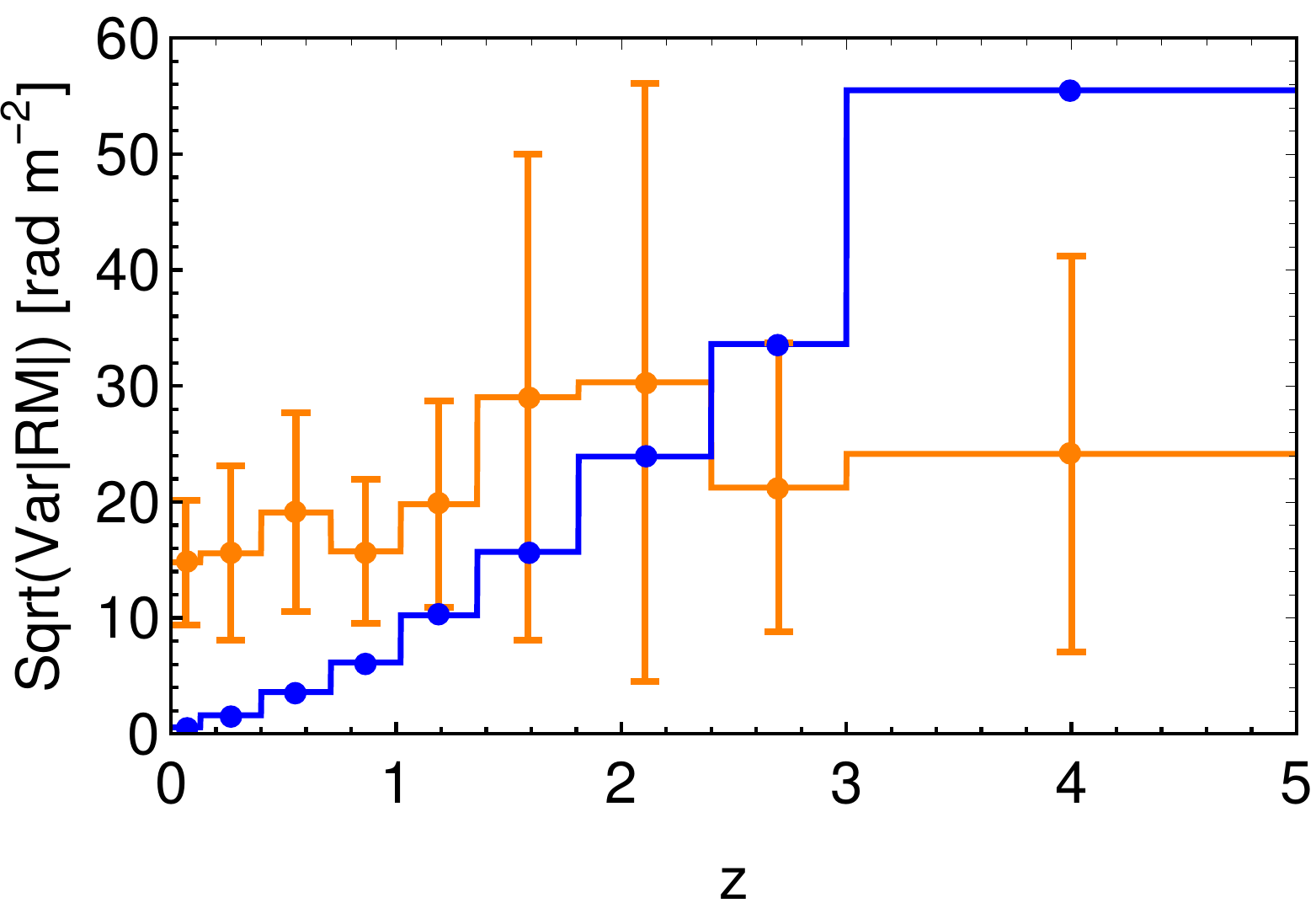}
    \caption{Example of expected dispersion for the limiting model of the primordial magnetic field with $B_0 = 1.8\cdot 10^{-8}$~cG (blue line). The orange line shows the corresponding dispersion of the observational data.}
    \label{fig:prim-var}
\end{figure}

In addition to the mean and medium values used in the main text, another statistic that is discussed in the literature to describe the extragalactic contribution to RM is a dispersion $\sigma_{\text{RM}}$, defined as the square root of the variance, see e.g.~\cite{Oppermann:2014cua}. For completeness, in this appendix, we show the results for this statistic as well.

Using the same methodology as discussed in Section~\ref{sec:constraints}, we calculate dispersion in the observational data in the same bins as in the main text, see Table~\ref{tab:zbins-new}, and make a prediction for the primordial magnetic field using 1000 lines of sight from the IllustrisTNG simulation. To calculate the error of the variance in the observational data, we use the formula~\citep{cho2005variance}:
\begin{equation}
    \Delta \text{Var}(x) = \frac{1}{n} \left( \mu_4 - \frac{n-3}{n-1} \text{Var}(x) \right),
\end{equation}
where $\mu_4$ is the fourth central moment, and $n$ is the number of data points in the bin. 
Following the same procedure as in Section~\ref{sec:results}, we place a constraint on the homogeneous primordial magnetic field from dispersion of $B_0 < 1.8\cdot 10^{-8}$~cG. An illustration of the $2\sigma$ exclusion for this limit is shown in Fig.~\ref{fig:prim-var}. This exclusion comes only from the two highest redshift bins, as the contribution to dispersion from $B_0$ can be smaller but not larger than that in the observed data. It can be seen that the constraint from this method is much weaker than that from either the mean or median |RM| values.
%

\bsp	
\label{lastpage}
\end{document}